\documentclass[twocolumn,amsmath,amssymb]{revtex4}

\usepackage{graphicx}
\usepackage{dcolumn}
\usepackage{bm}
\newcommand{\etal}{\emph{et al.}}
\newcommand{\be}{\begin{equation}}
\newcommand{\ee}{\end{equation}}
\newcommand{\bfig}{\begin{figure}}
\newcommand{\efig}{\end{figure}}
\newcommand{\incl}{\includegraphics}

\begin{document}      
\title{Reply to Comment
}

\date{\today}     
\begin{abstract}
\end{abstract}

\maketitle                   
Our experiments~\cite{Wang05,Li05,Wang06} persuade us that the Meissner 
transition at $T_c$ in hole-doped cuprates is driven by the loss of long-range 
phase coherence caused by singular phase fluctuations, 
a scenario at odds with the mean-field (MF), Gaussian Ginzburg Landau (GGL) approach 
advocated by Cabo, Mosquiera and Vidal~\cite{Cabo}.  

\bfig [h]  
\incl[width=8cm]{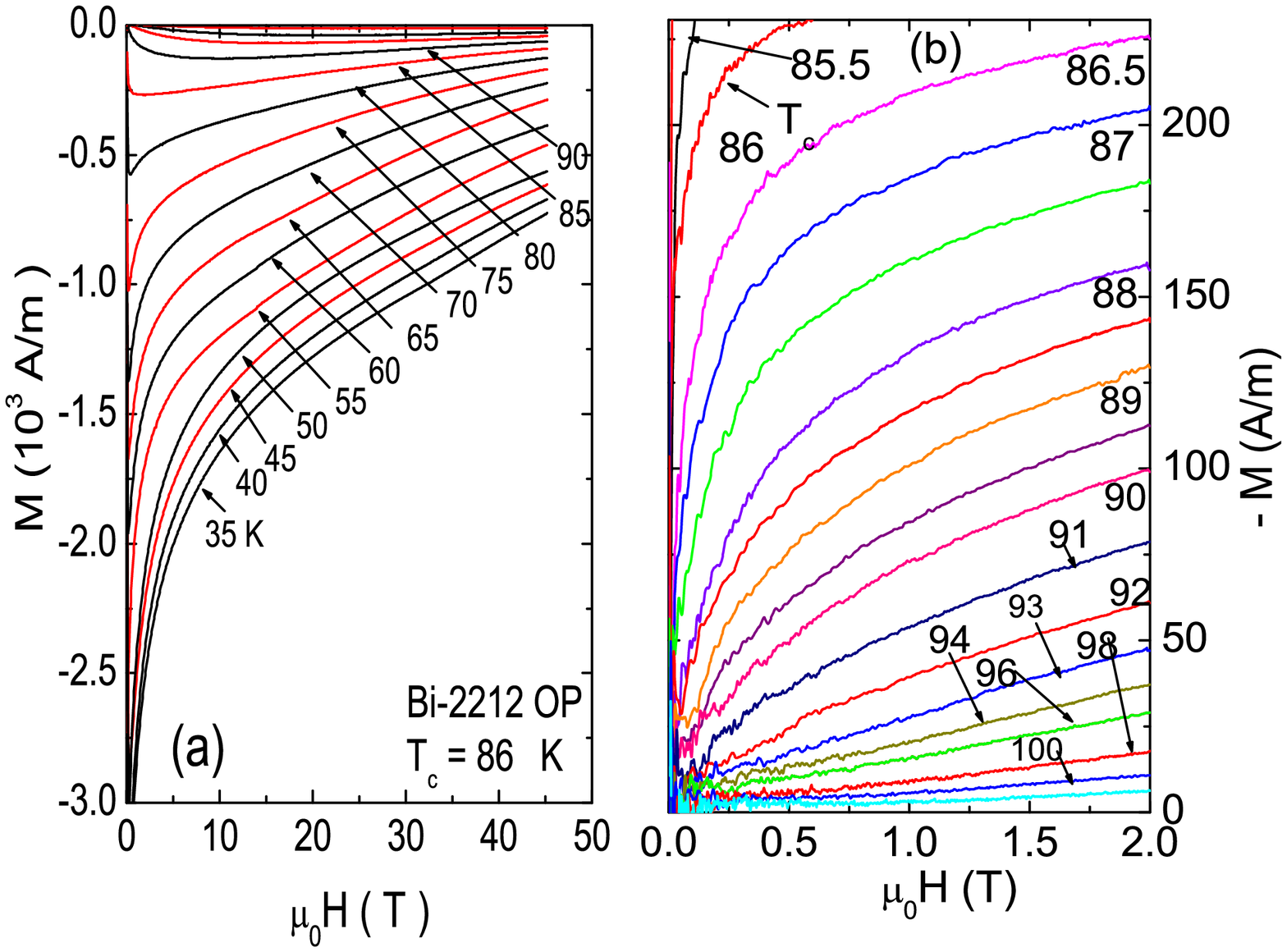}
\caption{\label{FigMH} (color online) 
Magnetization curves in OP Bi 2212.  Panel a shows $M$ measured 
up to 45 T at $T$ from 35 K to above 90 K ($T_c$ = 86 K).  Below 70 K, $|M|$ 
decreases as $\log H$ over a very broad field interval~\cite{Wang05}, but above 70 K, 
$M$ vs. $H$ shows anomalous features such as the separatrix $T_s$ ($\sim$85 K) at 
which $M$ is independent of $H$ below $\sim$5 T.  In large $H$, $|M|$ again decreases 
as $\log H$. $H_{c2}(T)$, defined as where $M\rightarrow 0$, remains high at 
values 100-150 T even when $T_c$ is exceeded, in sharp contrast to GGL (MF) 
predictions (previous experiments on $M$ stop at 5 T).  Panel b displays the strong 
curvature of $M$ above $T_c$ in weak $H$ ($<$2 T)~\cite{Li05}.  The curves 
display persistent negative curvature consistent 
with $M\sim -H^{1/\delta}$ with a $T$-dependent $\delta(T)$ up to $\sim$105 K.  
}
\efig
A characteristic MF feature is the linear decrease  to zero of the upper critical 
field $H_{c2}(T)\sim (1-t)$ near $T_c$ ($t=T/T_c$, with $T$ the temperature).
In sharp contrast, magnetization ($M$) 
and Nernst data in intense fields $H$ show that, in hole-doped cuprates, 
$H_{c2}(T)$ remains very large up to $T_c$~\cite{Wang05,Li05,Wang06}.   
Figure \ref{FigMH}a shows curves of $M$ in optimal (OP) 
$\rm Bi_2Sr_2CaCu_2O_{8+\delta}$ (Bi 2212) measured in fields $H$ to 45 T.  
The estimated $H_{c2}$ values remain significantly higher than 45 T as $T$ is raised above
86 K.  The curves directly contradict previous, inferred $H_{c2}\sim (1-t)$ behavior, mostly from data taken below 5 T (see Fig. 3 of Ref. \cite{Wang06} for curves in OP $\rm YBa_2Cu_3O_7$).

A second incompatibility with the GGL approach is the
striking nonlinearity of the $M$-$H$ curves in Bi 2212 (Fig. \ref{FigMH}b)~\cite{Li05}.
Between 105 K and $T_c$, the magnetization displays the fractional power-law behavior
$M\sim -H^{1/\delta}$ in weak fields~\cite{Li05}.  The exponent $\delta(T)$ grows 
from 1 to $\sim 15$ as $T$ falls from 105 K to $T_c$.  
This unusual field dependence is in conflict with the Gaussian treatment of fluctuations.
Other incompatibilities with GGL theory include the anomalous increase in $|M|$ 
above $H_{c1}$ for $T<T_c$ (see Ref. \cite{Li05} for full discussion).

In light of the fundamental incompatibilities, fitting the $T$ dependence of $M(T,H)$ at 
a \emph{single} value of $H$ is not an enlightening exercise.  As shown in 
Ref. \cite{Li05}, $|M(T,H)|$ with $H$ = 10 Oe actually diverges exponentially as 
$T$ decreases towards $T_c$ from above, instead of as a power-law in ($t-1$).
Approaches based on phase-disordering schemes, e.g.
the Kosterlitz-Thouless transition~\cite{Vadim} or the anisotropic 3D XY model~\cite{Sudbo}
seem more productive.

Further, the GGL fits lead to parameters that are highly unreliable.  
In optimal Bi 2212, the experiment gives $H_{c2}(0)$ = 150-200 T, 
whereas the fit in Ref.~\cite{Cabo} predicts 330 T.   In underdoped LSCO ($x$ = 0.1), we 
find $H_{c2}(0)\sim$ 80 T, whereas a similar analysis~\cite{Vidal} predicts 
26--28 T.  Such a low $H_{c2}$ is ruled out by experiment~\cite{Wang06,Li07}.  

Other evidence exist for the survival of the pair condensate amplitude high above $T_c$
in the pseudogap state.  These include the Nernst effect~\cite{Wang06}, kinetic 
inductance~\cite{Corson} and measurements of the gap~\cite{Renner,Yazdani}.  
These anomalies are completely outside the purview of Gaussian fluctuations that
underlie the GGL approach, regardless of the cut-off scheme adopted.
\vspace{3mm}\\
Lu Li$^1$, Yayu Wang$^1$, M. J. Naughton$^2$ and N. P. Ong$^1$\\
$^1$ Princeton University\\
$^2$ Boston College

\end{document}